%% file: main.tex
\documentclass{CUP-JNL-DTM}%

\usepackage{graphicx}
\usepackage{multicol,multirow}
\usepackage{amsmath,amssymb,amsfonts}
\usepackage{mathrsfs}
\usepackage{amsthm}
\usepackage{rotating}
\usepackage{appendix}
\usepackage{nomencl}
\usepackage{etoolbox}
\usepackage[numbers,sort&compress]{natbib}
\usepackage{ifpdf}
\usepackage[T1]{fontenc}
\usepackage{newtxtext}
\usepackage{newtxmath}
\usepackage{textcomp}
\usepackage{xcolor}
\usepackage{lipsum}
\usepackage[colorlinks,allcolors=blue]{hyperref}
\usepackage{array}

\usepackage{tikz}
\usetikzlibrary{shapes}
\usepackage{soul}
\usepackage{subcaption}
\usepackage{hyperref}
\hypersetup{
    colorlinks = true,
    urlcolor   = blue,
    citecolor  = black,
}
\usepackage{placeins}
\usepackage{mleftright}

\theoremstyle{definition}

\numberwithin{equation}{section}

\usepackage{listings}
\usepackage{algorithm2e}

\jname{}
\articletype{ARTICLE TYPE}
\jyear{YEAR}

\def\LW#1{\dimexpr#1\linewidth-.5em}

\def\mat#1{\pmb{#1}}
\def\vector#1{\boldsymbol{#1}}

\def\sym#1{\mbox{\textit{#1}}}
\def\loss{\mathcal{L}}
\def\Re{\mbox{\textit{Re}}}

\def\loss{\mathcal{L}}

\def\lossStrict{\loss^s}
\def\lossMean{\loss^m}

\begin{document}

\begin{Frontmatter}

\title[Article Title]{Reconstruction of three-dimensional turbulent flows from sparse and noisy planar measurements:\\ A weight-sharing neural network approach}

\author[1]{Yaxin Mo}
\author[1,2]{Luca Magri}
\authormark{Yaxin Mo \textit{et al}.}
\address[1]{
    \orgdiv{Department of Aeronautics}, 
    \orgname{Imperial College London}, 
    \orgaddress{\city{London}, \postcode{SW7 2AZ},  \country{UK}}
    \email{l.magri@imperial.ac.uk}
}

\address[2]{
    \orgname{Politecnico di Torino, DIMEAS}, 
    \orgaddress{\city{Torino}, \postcode{10129},  \country{Italy}}
}


\abstract{
This paper proposes a method for reconstructing three-dimensional turbulent flows from sparse measurements without the need for ground truth data during training. A weight-sharing network is developed to infer the full flow fields from measurements of  velocity sampled at three planes and boundary pressure at one additional plane, inspired by experimental configurations. The weight-sharing network shares identical parameters along homogeneous directions, which results in efficient data utilization and reduced computational memory requirements. First, we compare the weight-sharing network to the PC-DualConvNet, adapted from prior work, by reconstructing a 3D Kolmogorov flow from noise-free measurements with a snapshot-enforced loss. Both networks accurately recover time-averaged 3D flow fields and the correct energy spectrum up to wavenumber 10. The weight-sharing network has the ability to infer flow structures distant from measurement planes. Second, we carry out reconstruction from measurements corrupted with white noise (SNR 15) using a mean-enforced loss. We show that, for the weight-sharing network, validation sensor loss on unseen data decreases with training sensor loss—unlike PC-DualConvNet. This shows improved generalization and that training sensor loss estimates generalization error. The weight-sharing network offers good generalization, parameter efficiency, and hyperparameter robustness. The proposed method opens the possibility of three-dimensional flow reconstruction from experiments.
}

\end{Frontmatter}

\section*{Impact Statement}
The weight-sharing network enables efficient and robust reconstruction of three-dimensional turbulent flows from sparse measurements, without requiring ground truth data during training. This capability expands the potential for practical 3D flow reconstruction in experimental settings and advances the design of data-driven models for fluid mechanics systems.

\section{Introduction}\label{sec:flowrec3d:intro}

In many practical applications concerning turbulent flows, the data is often sparse. Various strategies have been developed to reconstruct flow fields from sparse measurements. Modal decomposition-based methods, such as gappy proper orthogonal decomposition (POD), have been used to estimate the flow in areas without measurements but may not work well when these areas are large \citep{everson1995KarhunenLoeveProcedure,venturi2004GappyDataReconstruction,gunes2006GappyDataKrig,nekkanti2023GappySpectralProper}. Sequential methods, such as the ensemble Kalman filter \citep{geirevensen2009DataAssimilationEnsemble}, can reconstruct flows from sensor measurements, but their accuracy depends on the quality of the incorporated reduced-order model \citep{colburn2011StateEstimationWallbounded,mons2016ReconstructionUnsteadyViscous,suzuki2017EstimationTurbulentChannel,novoa2024InferringUnknownUnknowns}. Adjoint-based variational methods can be used to accurately reconstruct flows from sensor measurements. Although variational methods constrain physical constraints, they may become unstable over a long reconstruction window in chaotic flows \citep{zaki2021LimitedObservationsState,mons2016ReconstructionUnsteadyViscous,franceschini2020MeanflowDataAssimilation,wang2014LeastSquaresShadowing,huhn2020stability}. Ensemble-variational methods have been used to reconstruct 3D unsteady channel flows from instantaneous sparse measurements and statistical observations \citep{mons2021EnsemblevariationalAssimilationStatistical,wang2021StateEstimationTurbulent,zaki2021LimitedObservationsState}, and to reconstruct mean flows from particle image velocimetry measurements \citep{he2025DataAssimilationNew}. These methods, however, require a large number of 3D simulations, which makes them computationally expensive.

Neural networks have also been increasingly applied to reconstruct velocity and pressure fields. From sparse measurements, network architectures such as autoencoders \citep{erichson2020ShallowNeuralNetworks,dubois2022MachineLearningFluid,kelshaw2024PhysicsconstrainedConvolutionalNeural}, generative adversarial networks \citep{buzzicotti2021ReconstructionTurbulentData,yousif2023DeeplearningApproachReconstructing,yu2022ThreedimensionalESRGANSuperresolution,kim2021UnsupervisedDeepLearning}, and transformers \citep{santos2023DevelopmentSenseiverEfficient} have been used to reconstruct two-dimensional (2D) flows including bluff body wakes, isotropic turbulence, and rotating turbulence. From 2D low-resolution data, convolutional neural networks (CNNs) are commonly used to recover high-resolution 2D flow fields \citep{fukami2019SuperresolutionReconstructionTurbulent,guemes2022SuperresolutionGenerativeAdversarial,matsuo2024ReconstructingThreeDimensionalBluff,kelshaw2024PhysicsconstrainedConvolutionalNeural}. Reconstructing three-dimensional (3D) turbulent flows poses a greater challenge because of the high dimensionality of the problem. \citet{chatzimanolakis2024LearningTwoDimensions} controlled 3D flows with a learning algorithm pre-trained on 2D flows. Physics-informed neural networks (PINNs) \citep{raissi2019PhysicsinformedNeuralNetworks} have been used to reconstruct 3D flows in different sensor setups, such as turbulent channel flow from tracked particles \citep{clarkdileoni2023ReconstructingTurbulentVelocity}, and wakes from sparse, point measurements \citep{zhang2021ThreedimensionalSpatiotemporalWind,rui2023Reconstruction3DFlow}. Convolutional neural networks (CNNs) \citep{ozbay2023FR3DThreedimensionalFlow,xuan2023ReconstructionThreedimensionalTurbulent} and generative models \citep{dubois2022MachineLearningFluid,yu2022ThreedimensionalESRGANSuperresolution,yousif2023DeeplearningApproachReconstructing} have been employed to infer missing information from limited measurements in 3D domains. Instantaneous velocities within wall-bound flows have been reconstructed from wall quantities only using CNNs and generative adversarial networks \citep{guastoni2021ConvolutionalnetworkModelsPredict,cuellar2024ThreedimensionalGenerativeAdversarial}. However, these methods require the full flow field to be known during training, which limits their applicability when the full flow field is not available.

When ground truth data is unavailable, physics-informed neural networks have been used to infer the flow field at unknown locations from known point measurements, but the training of PINNs requires knowing the time coordinates \citep{raissi2019PhysicsinformedNeuralNetworks,zhang2021ThreedimensionalSpatiotemporalWind}. Without initial conditions and with data organised on regular grids, computer vision approaches have been used to reconstruct 3D flows. CNNs have been used to reconstruct flows such as 2D steady biological flows \citep{gao2021SuperresolutionDenoisingFluid} and 2D turbulent Kolmogorov flows \citep{kelshaw2024PhysicsconstrainedConvolutionalNeural,mo2025ReconstructingUnsteadyFlows,page2025SuperresolutionTurbulenceDynamics}. In particular, \citet{page2025SuperresolutionTurbulenceDynamics} and \citet{buzzicotti2021ReconstructionTurbulentData} tested their methods in 2D flows. \citep{raissi2019PhysicsinformedNeuralNetworks,buzzicotti2021ReconstructionTurbulentData,mo2025ReconstructingUnsteadyFlows} reconstructed flows without the full flow field in the training, using sensor setups that may be difficult to implement in experiments. When the data comes from experiments, the types of data and the locations of measurement points are limited by the experimental techniques. Non-intrusive methods, such as particle tracking, are often preferred over intrusive methods, as they do not disturb the flow.

Currently, there are no non-intrusive methods to measure static pressure directly \citep{tavoularis2024MeasurementFluidMechanics}, which appears in the incompressible Navier-Stokes equations. Instead, pressure is often obtained at walls via pressure taps \citep{mckeon2007PressureMeasurementSystems,cal2010ExperimentalStudyHorizontally,rowand.brackston2017FeedbackControlThreeDimensional}, or solved for from well-resolved velocity fields using the incompressible Navier-Stokes equations \citep{probsting2013EstimationWallPressure,vanoudheusden2013PIVbasedPressureMeasurement,tavoularis2024MeasurementFluidMechanics}. Among non-intrusive methods for measuring velocities, particle image velocimetry (PIV) is a robust method when spatially-resolved velocity is required. PIV provides in-flow velocity measurements on a regular grid and is capable of measuring time-resolved velocities up to three dimensions in a plane \citep{tavoularis2024MeasurementFluidMechanics}. Volumetric velocity can be obtained by measuring multiple planes in the same experiment. For example, by scanning the measurement planes across the domain \citep{zhu2024NewInsightsExperimental}, measuring simultaneously multiple planes arranged either in parallel \citep{ganapathisubramani2005DualplanePIVTechnique,pfadler2009HighResolutionDualplane,cal2010ExperimentalStudyHorizontally} or perpendicular to each other \citep{morton2016ReconstructingThreedimensionalWake,chandramouli2019Fast3DFlow}, or using tomographic PIV \citep{probsting2013EstimationWallPressure}. The regular grid and spatial resolution of PIV measurements make them a good starting point for 3D flow reconstruction.

Using POD, \citet{druault2007UseProperOrthogonal} reconstructed the mean in-cylinder flow from multiple PIV planes; \citet{hamdi2018VolumeReconstructionImpinging} reconstructed an impinging jet from multiple parallel planes; and \citet{chandramouli2019Fast3DFlow} reconstructed a 3D turbulent flow from two planes perpendicular to each other, using both experimental and synthetic data. A 3D stratified flow has also been reconstructed from multiple experimental PIV planes using PINN \citep{zhu2024NewInsightsExperimental}. Synthetic data on a regular grid have also been used to develop methods for reconstructing 3D flows. \citet{perez2020ReconstructionThreedimensionalFlow} reconstructed a cylinder wake from multiple parallel planes of velocities. CNN-based methods have been used to reconstruct 3D free surface flow from surface measurements \citep{xuan2023ReconstructionThreedimensionalTurbulent}, and other turbulent flows from a cross-plane setup (two planes perpendicular to each other) \citep{yousif2023DeeplearningApproachReconstructing}. \citet{ozbay2023FR3DThreedimensionalFlow} reconstructed the 3D wake of a cylinder from both a cross-plane and multiple parallel planes taken from simulations.

Many physical systems such as flows have symmetries and obey conservation laws, which have been exploited in multiple works on flow reconstruction. One of these symmetries is homogeneity, i.e., the flow is statistically invariant under translation \citep{pope2000TurbulentFlows}. When reconstructing 3D flows from multiple planes, \citet{chandramouli2019Fast3DFlow} used the homogeneous assumption to design their reconstruction method so that the method does not require the full 3D flow field. Neural networks can also be designed to enforce certain properties on their output, such as conservation of energy \citep{greydanus2019HamiltonianNeuralNetworks}, conservation of mass \citep{mohan2023EmbeddingHardPhysical,page2025SuperresolutionTurbulenceDynamics}, and periodicity \citep{ozan2023HardconstrainedNeuralNetworks}.

The overarching goal of this paper is to develop a neural network to reconstruct three-dimensional turbulent flows from sparse and noisy data, inspired by experimental configurations, without using the full flow field in the training. The method is tested on the flow reconstruction of 3D turbulent flows from a small number of planes of velocity measurements and a plane of pressure measurements at the boundary of the flow. We design a CNN-based weight-sharing network to both reduce the number of parameters needed for the reconstruction of the 3D flow and to exploit the homogeneous directions in the flow. The paper is structured as follows. We describe the sensor setup and the network in Section~\ref{sec:flowrec3d:methods}. We present the reconstructed flow from non-noisy measurements in Section~\ref{sec:flowrec3d:results-clean}, and from noisy measurements in Section~\ref{sec:flowrec3d:results-noisy}. We present our conclusion in Section~\ref{sec:flowrec3d:conclusion}.

\section{Methodology}\label{sec:flowrec3d:methods}
In this section, we first describe the dataset to be reconstructed and how the measurements are taken (Section~\ref{sec:flowrec3d:methods-data}).
Then, we introduce the network designed for 3D reconstruction (Section~\ref{sec:flowrec3d:methods-nn}).

\subsection{Dataset and measurements}\label{sec:flowrec3d:methods-data}

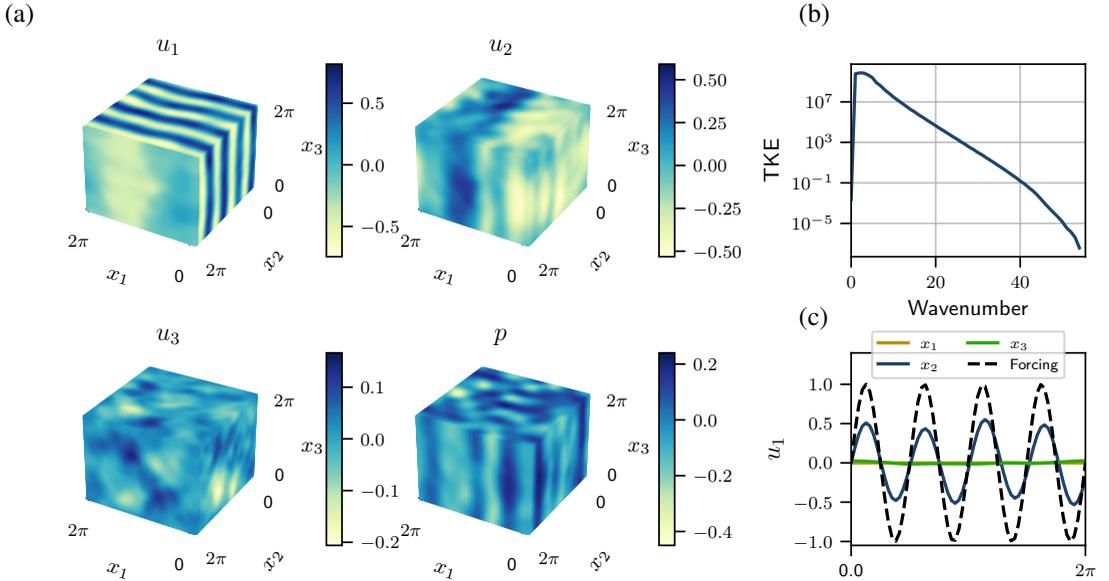
\begin{figure}
    \centering
    \begin{tikzpicture}
        \node[inner sep=0pt] (img) at (0,0) {\includegraphics[width=\linewidth]{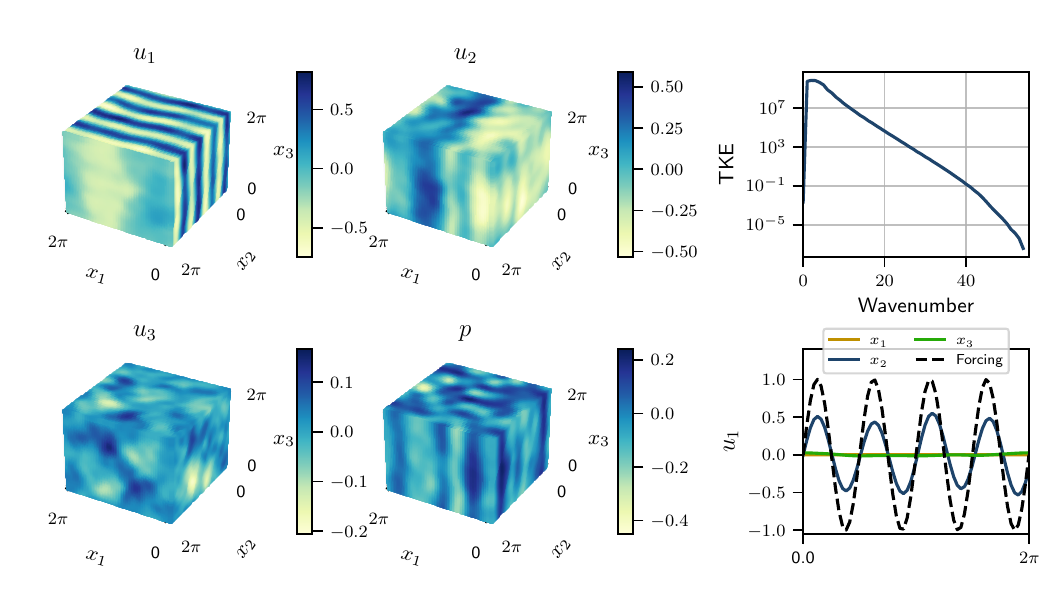}};
        \node[anchor=south west] at (-7.5,3.5) {(a)};
        \node[anchor=south west] at (3,3.5) {(b)};
        \node[anchor=south west] at (3,-0.5) {(c)};
    \end{tikzpicture}
    \caption{
        The mean properties of the 3D Kolmogorov flow dataset. 
        (a) Mean velocities and pressure. 
        (b) Turbulent kinetic energy spectrum. 
        (c) Mean $u_1$ averaged over each of the spatial directions, compared with the forcing term.  
    }\label{fig:extend:method-3dkol-data}
\end{figure}

The 3D turbulent Kolmogorov flow dataset is generated with a pseudo-spectral solver KolSol \citep{kelshawdaniel2023MagriLabKolSolPseudospectral} by solving the incompressible Navier-Stokes equations
\begin{equation}
    \begin{cases}
        \nabla \cdot \vector{u} = \mathcal{R}_{d} (\vector{u})\\
        \frac{\partial\vector{u}}{\partial t} + \vector{u} \cdot \nabla \vector{u} + \nabla p - \frac{1}{Re}\triangle\vector{u} - \vector{g}= \mathcal{R}_{m}(\vector{u},p) ,
    \end{cases}\label{eq:ns-equations}
\end{equation}
where $\vector{u}(\vector{x},t) \in \mathbb{R}^{N_u}$ and $p(\vector{x}, t) \in \mathbb{R}$ are the velocity and pressure at location $\vector{x}$ and time $t$, $N_u$ is the number of velocity components, and $\mathcal{R}_{(\cdot)}$ is a residual of the equation, which is zero when the equation is exactly solved. 
With this non-dimensionalization, the Reynolds number $\sym{Re}$ is the inverse of the kinematic viscosity.
The flow is subjected to sinusoidal forcing $\vector{g}=\vector{e}_1\sin(k\vector{x})$, where $\vector{e}$ is a standard unit vector.
The dataset, $\mat{D}$, consists of four time series of 3D Kolmogorov flows, initialised with different initial conditions.
Each time series is generated with $\Re=32$, with 32 wavenumbers and the time step $\Delta t^* = 0.005$.
A snapshot is saved every 20 time steps and interpolated onto a grid of $64 \times 64 \times 64$ points in the physical space, resulting in a time step of $\Delta t = 0.1$.
Each time series contains 500 snapshots, which is longer than the decorrelation time of the Kolmogorov flow. 
The combined dataset $\mat{D}$ contains $2000$ snapshots.
The dataset is validated by comparing its time-averaged properties with those found in the literature.
The mean $u_1$ averaged across three directions is shown in Figure~\ref{fig:extend:method-3dkol-data}c, where the average $u_1$ in the forced direction $x_2$ has a sinusoidal profile matching the frequency of the forcing term, and the averaged velocity is $0$ in other directions.
Given a long enough dataset, the mean of a 3D turbulent Kolmogorov flow is expected to have the same velocity profile as its laminar form, independent of the Reynolds number \citep{borue1996NumericalStudyThreedimensional}, which is  shown in the 3D plots of velocities and pressure in Figure~\ref{fig:extend:method-3dkol-data}a.
The turbulent kinetic energy spectrum shows the exponential decay of energy, which matches the results obtained by \citet{shebalin1997KolmogorovFlowThree}.
Figure~\ref{fig:extend:method-3dkol-data} shows that the combined dataset has converged and has the expected mean velocity profile and energy spectrum.

\begin{figure}
    \centering
    \includegraphics[width=3in]{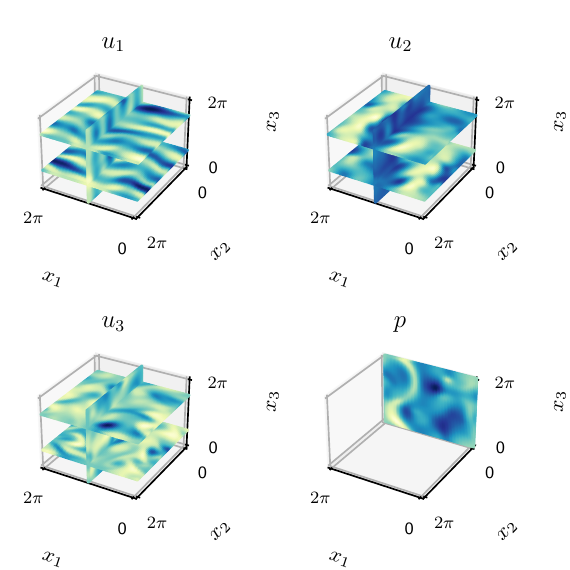}
    \caption{Pressure is measured at the plane $x_2=0$. Three planes of velocities are taken at $x_1=3.14$, $x_3=1.57$, and $4.71$. The pressure is used as input to the network, and all measurements are used as collocation points.}\label{fig:flowrec3d:methods-data}
\end{figure}
All three velocity components are measured at all grid points $\vector{x}_s$ on three planes: an $x_2 - x_3$ plane at $x_1=3.14$, and two $x_1 - x_2$ planes at $x_3=1.57$ and $4.71$. 
Pressure is measured at all grid points $\vector{x}_{in}$ on the $x_1 - x_3$ plane at $x_2=0$.
Velocities and pressure are measured at different planes to reflect that these quantities are typically measured with  different instruments in experiments.
The planes are shown in Figure~\ref{fig:flowrec3d:methods-data}
The collection of all measurements $\xi(\mat{D}) = \{\mat{U}(\vector{x}_s),\mat{P}(\vector{x}_{in})\}$ contains both the velocity measured at $\vector{x}_s$ and the pressure measured at $\vector{x}_{in}$.
The pressure measurements $\mat{P}(\vector{x}_{in})$ are used as inputs to the network.
The measurements account for approximately 3.8\% of the total number of variables in a snapshot.
The number of $x_1 - x_2$ planes is determined by testing and reducing the number of planes until the relative errors from both networks exceed 50\%; the details can be found in Appendix~\ref{sec:app-ml:chap-flowrec3d-params-and-min-planes}.

\subsection{The physics-constrained dual-branch convolutional neural network}\label{sec:methods:pc-dualconvnet}
\begin{figure}
    \centering
    \includegraphics[width=0.9\linewidth]{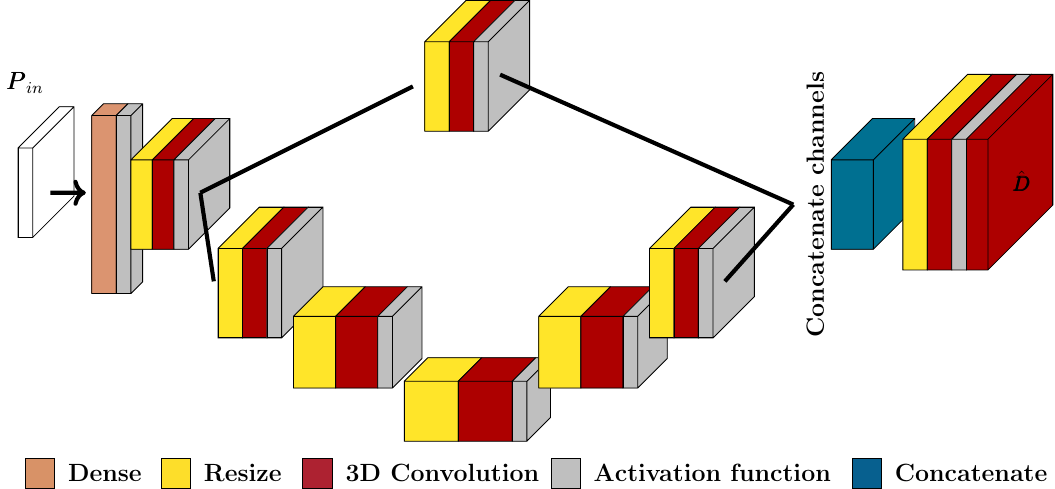}
    \caption{Schematic of the PC-DualConvNet with 3D convolutions.}\label{fig:methods-pc-dualconvnet}
\end{figure}

In previous work, we designed a physics-constrained dual-branch convolutional neural network (PC-DualConvNet) to reconstruct 2D flows from sparse measurements in \citep{mo2025ReconstructingUnsteadyFlows}. 
In this paper, we develop a scalable PC-DualConvNet to reconstruct 3D flows (Figure~\ref{fig:methods-pc-dualconvnet}), which we will now refer to as the PC-DualConvNet.
Periodic padding is used in convolutions to reflect the periodic boundary conditions of the flow under investigation.
We will refer to the 2D version of the PC-DualConvNet used in \citet{mo2025ReconstructingUnsteadyFlows}, which constitutes  part of the weight-sharing network (Section~\ref{sec:flowrec3d:methods-nn}), as the 2D PC-DualConvNet.

\subsection{The weight-sharing network}\label{sec:flowrec3d:methods-nn}
Part of the difficulty in reconstructing 3D flows is the large demand on computational resources.
CNNs need fewer parameters than other types of commonly used networks, such as fully-connected networks or transformers, but a large amount of resources is still required for three-dimensional convolutions.
The Kolmogorov flow is statistically homogeneous in all but the forced direction \citep{borue1996NumericalStudyThreedimensional}.
When homogeneous directions are present, the statistical dimension of the flow is reduced \citep{pope2000TurbulentFlows}.
For example, if one direction is homogeneous in a 3D flow, then the flow is statistically 2D.
We develop a weight-sharing network to both reduce the number of parameters and to fully utilise the homogeneity in the  flow. 
Part of the network parameters are shared across the $x_3$ direction, hence the name weight-sharing network.
The shared part is based on the Physics-constrained Dual-branch Convolutional Neural Network (PC-DualConvNet) \citep{mo2025ReconstructingUnsteadyFlows}, which is the 2D version of the PC-DualConvNet presented in Section~\ref{sec:methods:pc-dualconvnet}.

\begin{figure}[htb]
    \centering
    \includegraphics[width=4.5in]{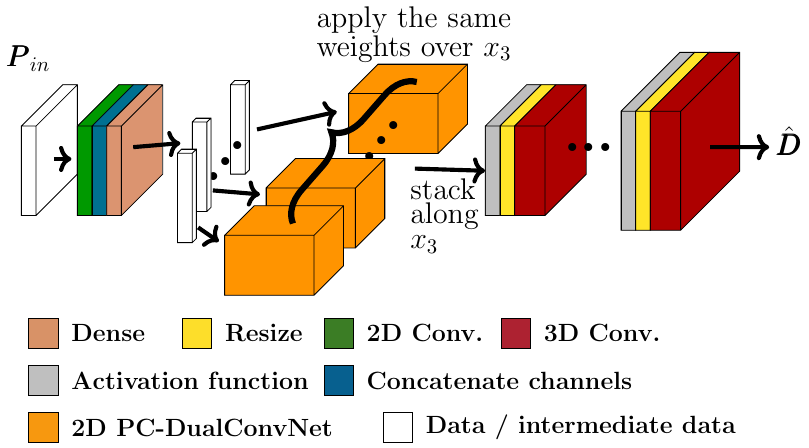}
    \caption{
        Schematics of the weight-sharing network.
        The input pressure $\mat{P}_{in}$ (first white block) is a 2D matrix of pressure measurements at $x_2=0$.
        It is passed through a 2D convolution layer and a fully-connected layer to produce another 2D matrix, which is then split into vectors along an axis, which will later become $x_3$ direction in the final output.
        Each vector is then passed through the same PC-DualConvNet (orange block) to produce an intermediate 2D result representing a $x_1-x_2$ plane.
        These intermediate results are stacked along $x_3$ direction to form a 3D intermediate result, which is then passed through multiple 3D convolutions and reshaping layers to produce the final reconstructed flow $\hat{\mat{D}}$.
    }\label{fig:flowrec3d:methods-network}
\end{figure}
The network, shown in Figure~\ref{fig:flowrec3d:methods-network}, can be broken down into three parts:
\begin{itemize}
    \item \textbf{Input processing:} Pressure input $\mat{P}_{in}$ at any instance in time, which is a 2D matrix, is passed through a 2D convolutional layer and a fully-connected layer, resulting in a 2D matrix.
    \item \textbf{The 2D inner network:} The 2D matrix from the previous step is split along an axis, which will  become $x_3$ in the output, into vectors. Each vector is passed through the same PC-DualConvNet (orange block) and becomes an intermediate result on a $x_1-x_2$ plane, to enforce that  homogeneous directions  are statistically invariant. These intermediate planes are then stacked along the $x_3$ direction.
    \item \textbf{The 3D CNN:} The stacked output from the previous step is passed through multiple layers of 3D convolutions and resizing via linear interpolation to produce the final output with the correct dimensions.
\end{itemize}
If we had an infinite number of snapshots of a single $x_1-x_2$ plane, then every possible realisation of the flow on a plane would be represented in the training data.
However, when only a finite number of snapshots is available, only parts, and not all, of the weights are shared across the $x_3$ direction to avoid too much restriction on the network.
Sharing parts of the weights informs the network that the flow is statistically similar along the $x_3$ direction, which is an  efficient use of available data.

\subsection{Mean-enforced loss and snapshot-enforced loss}

Neural networks are trained to minimise the value of a loss function $\loss$, which measures the error between the reference data $\mat{D}$ and the reconstructed flow $\hat{\mat{D}}$.
The loss function contains information on both the measurements and the physics of the flow.
We define the sensor loss $\loss_o$ to be 
\begin{equation}
    \loss_o (\hat{\mat{D}}, \mat{D}) = \| \xi(\hat{\mat{D}}) - \xi(\mat{D}) \|^2_2,
    \label{eq:method:sensor-loss}
\end{equation}
which is the $\ell_2$ norm of the difference between the measurements and the reconstructed flow at the measurement planes.
We also define the physics losses
\begin{gather}
    \loss_{div}(\mat{D}) = \| \mathcal{R}_d(\mat{U}) \|^2_2, \\
    \loss_{mom}(\mat{D}) = \| \mathcal{R}_m(\mat{U,P}) \|^2_2,
    \label{eq:method:residual-loss}
\end{gather}
where $\loss_{mom}$ and $\loss_{div}$ are the $\ell_2$ norm of the residuals the momentum and continuity equations in~\eqref{eq:ns-equations}, respectively.

When the measurements are accurate and precise, the snapshot-enforced loss $\lossStrict$ \citep{gao2021SuperresolutionDenoisingFluid,mo2025ReconstructingUnsteadyFlows} minimises only the physics-related losses while the sensor loss is enforced to be $0$. 
The harder constraint on the sensor measurements means that the network cannot output trivial solutions.
We define the snapshot-enforced loss as
\begin{equation}
    \lossStrict = \lambda_{div} \loss_{div}(\mat{\Phi}) + \lambda_{mom} \loss_{mom}(\mat{\Phi}),
    \label{eq:strictly-enforced-loss}
\end{equation}
where $\mat{\Phi}^T = [\mat{\Phi}_u^T, \mat{\Phi}_p^T]$ is defined as 

\begin{minipage}{0.47\linewidth} 
    \begin{align*}
        \mat{\Phi_u}(\vector{x}) =
        \begin{cases}
            \mat{U}(\vector{x}) & \text{where } \vector{x} \in \mat{x}_s, \\
            \hat{\mat{U}}(\vector{x}) & \text{otherwise}.
        \end{cases}
    \end{align*}
\end{minipage}%
\begin{minipage}{0.5\linewidth} 
    \begin{align}
        \mat{\Phi_p}(\vector{x}) =
        \begin{cases}
            \mat{P}(\vector{x}) & \text{where } \vector{x} \in \mat{x}_{in}, \\
            \hat{\mat{P}}(\vector{x}) & \text{otherwise}.
        \end{cases}
    \end{align}
\end{minipage}
Practically, we enforce the measurements by replacing the network output with measurements if measurements are available for the grid points.
When the measurements are noisy, we use the mean-enforced loss $\lossMean$ \citep{mo2025ReconstructingUnsteadyFlows}, which are designed to reconstruct flows from measurements with white noise.
The mean-enforced loss enforces the mean of the measurements while placing a constraint on the instantaneous measurements.
The mean-enforced loss is defined as 
\begin{equation}
    \lossMean = \lambda_o \loss_o(\hat{\mat{D}}, \mat{D}) + \lambda_{div} \loss_{div}(\mat{\Phi}) + \lambda_{mom} \loss_{mom}(\mat{\Phi}),
    \label{eq:mean-loss}
\end{equation}
where $\mat{\Phi}^T = [\mat{\Phi}_u^T, \mat{\Phi}_p^T]$ is 

\begin{minipage}{0.47\linewidth} 
    \begin{align*}
        \mat{\Phi_u}(\vector{x}) =
        \begin{cases}
            \overline{\mat{U}}(\vector{x}) + \hat{\mat{U}}^\prime(\vector{x}) \; \text{where} \; \vector{x} \in \mat{x}_s, \\
            \hat{\mat{U}} \; \text{otherwise}.
        \end{cases}
    \end{align*}
\end{minipage}%
\begin{minipage}{0.5\linewidth} 
    \begin{align}
        \mat{\Phi_p}(\vector{x}) =
        \begin{cases}
            \overline{\mat{P}}(\vector{x}) + \hat{\mat{P}}^\prime(\vector{x}) \; \text{where} \; \vector{x} \in \mat{x}_{in}, \\
            \hat{\mat{P}} \; \text{otherwise}.
        \end{cases}
    \end{align}
\end{minipage}
The symbols $\overline{*}$ and $*'$ denote the time-averaged and fluctuating quantities, respectively.
For a more detailed explanation of the snapshot-enforced and the mean-enforced losses, the reader is referred to \citep{mo2025ReconstructingUnsteadyFlows}.

\section{Flow reconstruction from planes}\label{sec:flowrec3d:results-clean}
We show the results on the flow reconstruction of 3D turbulent flows from measurement planes and the snapshot-enforced loss.
We compare the results obtained using PC-DualConvNet and the weight-sharing network.
The network parameters for this section are listed in Appendix~\ref{sec:app-ml:chap-flowrec3d-params-and-min-planes}.

\begin{table}
    \centering
    \caption{
        The relative error $\epsilon$, the physics loss ($\loss_p$=$\loss_{mom}$+$\loss_{div}$), and the sensor loss $\loss_o$ (\textit{mean $\pm$ standard deviation}) of the reconstruction results from planes of measurements, averaged over five tests with different random initialisations of network weights.
    }\label{tab:flowrec3d:results-clean}
    \begin{tabular}{|c|c|c|c|}
        \hline
         & \textbf{$\epsilon$ (\%)} & \textbf{$\loss_p$} & \textbf{$\loss_o$} \\
        \hline
        \textbf{Reference data} & N/A & 0.137$\pm$0.000 & N/A \\
        \textbf{Weight-sharing network} & 49.3$\pm$0.4 & 0.324$\pm$0.092 & 0.0182$\pm$0.0011\\
        \textbf{PC-DualConvNet} & 54.7$\pm$7.4 & 0.354$\pm$0.109 & 0.0070$\pm$0.0009 \\
        \hline
    \end{tabular}
\end{table}
A summary of the results is shown in Table~\ref{tab:flowrec3d:results-clean}.
The weight-sharing network achieves lower values in both the relative error and the physics loss, and with a smaller standard deviation, compared to the PC-DualConvNet.
The relative error of a reconstructed flow $\epsilon$ is defined as
\begin{equation}
    \epsilon  = \sqrt{\frac{\| \hat{\mat{D}} - \mat{D} \|^2_2}{\| \mat{D} \|^2_2}}\;\;\;\;(\%).\label{eq:rel-l2}
\end{equation}
The physics loss $\loss_p$ is the unweighted sum of all physics-related losses, $\loss_{mom}$ and $\loss_{div}$.
The sensor loss of the weight-sharing network is over twice as large as that of the PC-DualConvNet, despite the weight-sharing network achieving a lower relative error and physics loss.
As the sensor loss measures only the data points on the measurement planes, we can see that the PC-DualConvNet overfits  the measurement planes.
By partially sharing weights across the $x_3$ direction, the weight-sharing network learns that all $x_1 - x_2$ planes are statistically similar, thereby reducing overfitting to the measurement planes.

\begin{figure}[]
\centering
\begin{subfigure}{0.9\linewidth}
    \includegraphics[width=\textwidth]{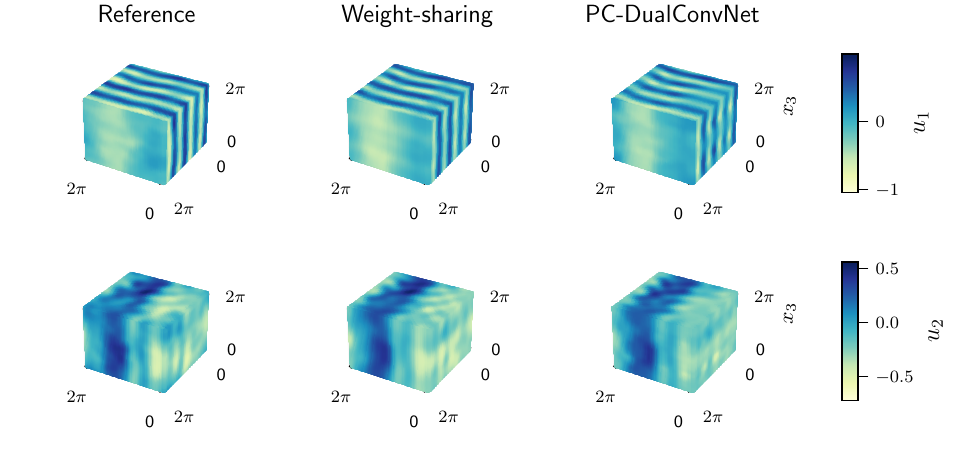}
\end{subfigure}
\begin{subfigure}{0.9\linewidth}
    \includegraphics[width=\textwidth]{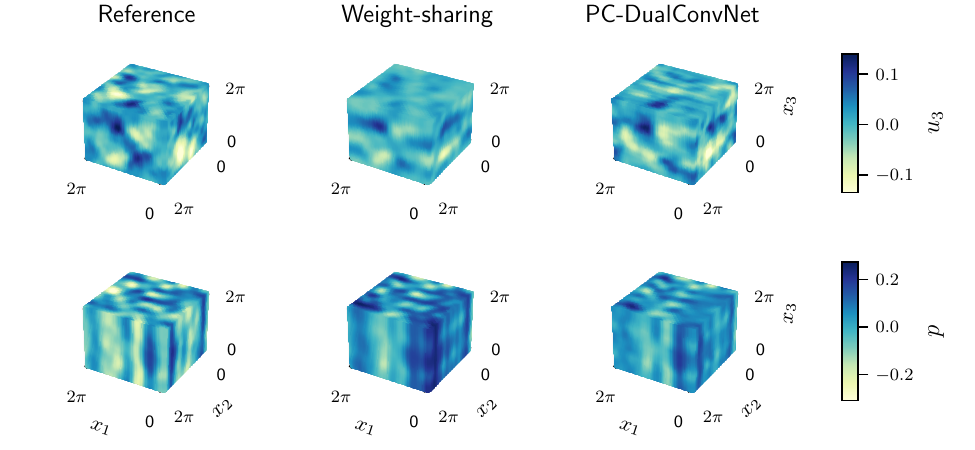}
\end{subfigure}
\caption{
    Time average of the volumes. From left to right, the columns are: the reference data, the flow reconstruction from the weight-sharing network, and the flow reconstruction from the PC-DualConvNet.
}\label{fig:flowrec3d:results-clean-mean-volume}
\end{figure}
\begin{figure}[]
    \centering
    \includegraphics[width=0.4\linewidth]{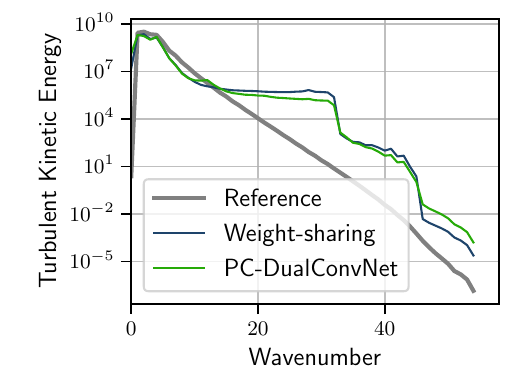}
    \caption{Energy spectrum.}\label{fig:flowrec3d:results-clean-tke}
\end{figure}
The results show good agreement with the reference data for both networks statistically.
There is little difference between the two networks when comparing the mean flow (Figure~\ref{fig:flowrec3d:results-clean-mean-volume}) or the energy spectrum (Figure~\ref{fig:flowrec3d:results-clean-tke}).
The networks correctly infer the correct energy spectrum up to approximately wavenumber 10, which contains the majority of the energy in the flow.

\begin{figure}
    \centering
    \begin{tikzpicture}
        \node[inner sep=0pt] (img) at (0,0) {\includegraphics[width=\linewidth]{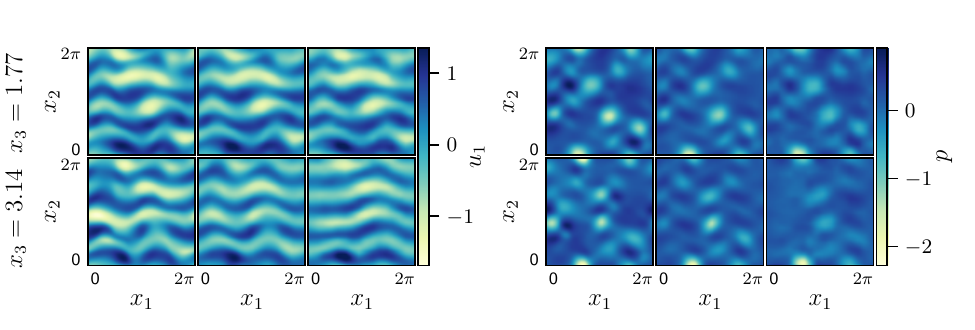}};
        \node[anchor=south west] at (-5.5,2) {\small{Ref.}};
        \node[anchor=south west] at (-4.2,2.2) {\small{Weight-}};
        \node[anchor=south west] at (-4.2,1.9) {\small{sharing}};
        \node[anchor=south west] at (-3,2) {\small{PC-DualConvNet}};
        \node[anchor=south west] at (1.2,2) {\small{Ref.}};
        \node[anchor=south west] at (2.5,2.2) {\small{Weight-}};
        \node[anchor=south west] at (2.5,1.9) {\small{sharing}};
        \node[anchor=south west] at (3.8,2) {\small{PC-DualConvNet}};
    \end{tikzpicture}
    \caption{
        Two $x_1-x_2$ planes taken from the reconstructed flow at $x_3=$1.77 and 3.14, which are unseen by the network during training. 
        The measurements used in training are taken on planes at $x_3 = $1.57 and 4.71. 
        Moving further away from the measured plane, the reconstruction from the weight-sharing network is closer  to the reference solution, whereas the PC-DualConvNet tends to converge toward the mean flow.
        An example of this difference can be seen in the bottom row, which shows a plane taken at $x_3 = $3.14. 
    }\label{fig:flowrec3d:results-clean-different-z}
\end{figure}
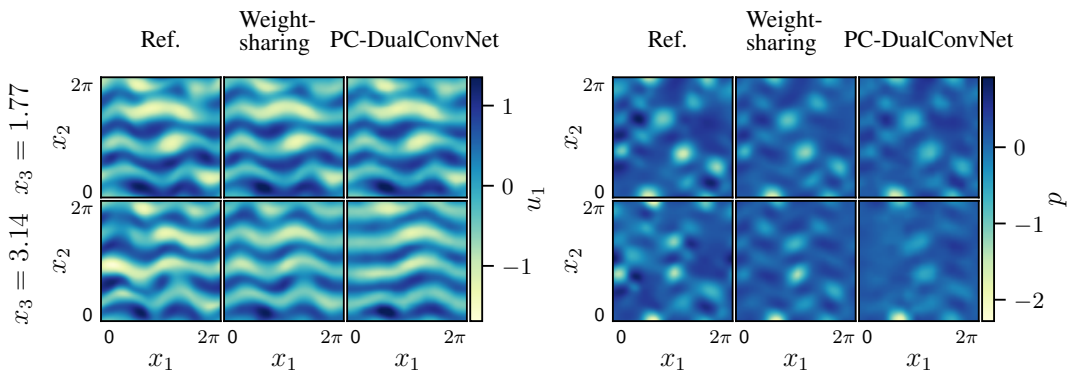
The differences between the networks are more visible when comparing individual planes within the 3D domain. 
Figure~\ref{fig:flowrec3d:results-clean-different-z} shows two instantaneous $x_1-x_2$ planes taken from the reference and reconstructed flow, the top row at $x_3=$1.77, which is close to the measured plane at $x_3=$1.57, and the bottom row at $x_3=3.14$, which is further away from any measured plane.
Both planes shown in Figure~\ref{fig:flowrec3d:results-clean-different-z} are unseen by the network during training. 
At both $x_3$, the reconstructed velocity $u_1$ (Figure~\ref{fig:flowrec3d:results-clean-different-z} left) from the weight-sharing network has retained the flow structures expected of an instantaneous snapshot.
However, PC-DualConvNet, which needs more parameters than the weight-sharing network, has larger errors at $ x_3=3.14$, and tends to converge toward the mean flow.
The weight-sharing network infers the  pressure field more accurately than the PC-DualConvNet, despite no pressure data has been used in training  (Figure~\ref{fig:flowrec3d:results-clean-different-z} right).

\subsection{Reconstructing from a single cross-plane}\label{sec:flowrec3d:single-crossplane}
\begin{figure}[htb]
    \centering
    \begin{subfigure}{0.4\linewidth}
        \includegraphics[width=\textwidth]{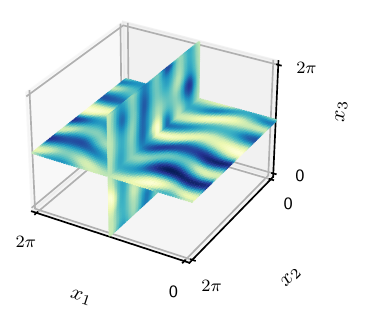}
        \caption{}\label{fig:flowrec3d:crossplane-sensors}
    \end{subfigure}
    \begin{subfigure}{0.4\linewidth}
        \includegraphics[width=\textwidth]{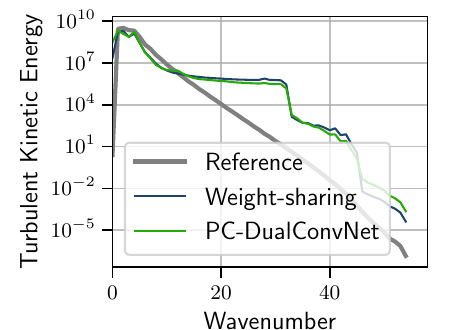}
        \caption{}\label{fig:flowrec3d:crossplane-tke}
    \end{subfigure}
    \begin{subfigure}{\textwidth}
        \includegraphics[width=\textwidth]{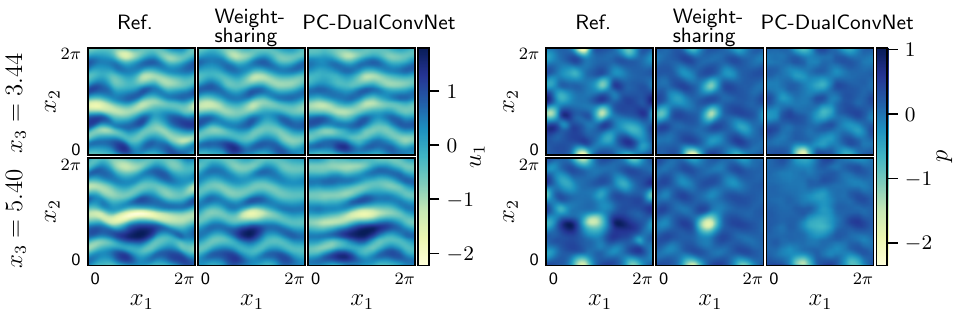}
        \caption{}\label{fig:flowrec3d:crossplane-slice}
    \end{subfigure}
    \caption{
        Reconstructed flow from a single cross-plane.
        (a) The locations of the velocity measurement planes, consisting of one $x_1-x_2$ measurement plane and a one $x_2-x_3$ measurement plane, at the centre of the domain.
        (b) The turbulent kinetic energy of the reconstructed flow fields.
        (c) 
        Two slices of the reconstructed flow at $x_3=$3.44 and 5.4, which are unseen by the network during training. 
        The measurements used in training are taken on planes at $x_3 = $3.14. 
    }\label{fig:flowrec3d:single-crossplane}
\end{figure}

In this section, we test the reconstruction from a single cross-plane.
Figure~\ref{fig:flowrec3d:crossplane-sensors} shows the location of the velocity measurements. 
The pressure inputs are taken from the same grid points as in Section~\ref{sec:flowrec3d:methods-data}.
Appendix~\ref{sec:app-ml:chap-flowrec3d-params-and-min-planes} provides more details on the tests to determine the minimum number of planes needed for accurate reconstruction.
When only a single cross-plane is used, the reconstruction relative errors for PC-DualConvNet and the weight-sharing network are 78\% and 62\%, respectively.
Both networks achieve a similar reconstructed turbulent kinetic energy (Figure~\ref{fig:flowrec3d:crossplane-tke}), which are not significantly different from reconstructing from two $x_1-x_2$ planes in Section~\ref{sec:flowrec3d:results-clean}.
By comparing slices in the reconstructed domain (Figure~\ref{fig:flowrec3d:crossplane-slice}), we can see that the reconstructed flow field by the PC-DualConvNet has lost resemblance to the reference data at $x_3=$5.4.
The difference between the networks is more pronounced in the pressure field, where the PC-DualConvNet fails to reconstruct the low pressure region in the centre of the slices, while the weight-sharing network successfully reconstructs those regions.

\section{Flow reconstruction  from noisy measurements}\label{sec:flowrec3d:results-noisy}
In this section, we reconstruct the flow using the same sensor setup as described in Section~\ref{sec:flowrec3d:methods-data}, but with added white noise.
The noise $e$ at a single instance in time and any grid point is drawn from a Gaussian distribution $e \sim \mathcal{N}(0,\sigma_e)$, where $\sigma_e$ is the standard deviation of the noise.
The signal-to-noise ratio (SNR) is defined as SNR$=10\log\left( \sigma^2 / \sigma^2_e \right)$, where $\sigma$ is the standard deviation of a component (such as a velocity or pressure) of the measurements.
In this section, we reconstruct the flow from measurements with an SNR=15, using the mean-enforced loss.
We show the process of selecting the hyperparameters in Section~\ref{sec:flowrec3d:results-noisy-params}, and the reconstructed flow in Section~\ref{sec:flowrec3d:results-noisy-results}.

\subsection{Selecting the hyperparameters}\label{sec:flowrec3d:results-noisy-params}

Before we can use the network to reconstruct the flow, we select a set of hyperparameters for the network that we expect will lead to an accurate reconstruction of the entire 3D flow.
We use only the noisy data in the hyperparameter selection. 
During the selection process, we perform multiple tests with the training dataset composed of the measurements taken from the planes shown in Section~\ref{sec:flowrec3d:methods-data} with added white noise at SNR=15.
The same noisy training dataset will also be used later in Section~\ref{sec:flowrec3d:results-noisy-results}.
Given that we are interested in a spatial reconstruction from sparse measurements, the validation dataset should provide information about a network's ability to generalise to unseen regions of the flow.
Thus, our validation dataset is the set of measurements from  $x_1-x_2$ plane at $x_3=$3.14, which is unseen by the network during training.
The validation sensor loss is then $\| \hat{\mat{U}}(x_3=3.14), \mat{U}_n(x_3=3.14) \|^2_2$, where $\hat{\mat{U}}$ is the reconstructed velocity field and $\mat{U}_n$ is the noisy reference velocity field.
The training sensor loss is $\|\xi(\hat{\mat{D}}), \xi(\mat{D}_n) \|^2_2$.
Since the training sensor loss is computed with pressure measurements, but the validation sensor loss is not, the two losses are not expected to be similar in magnitude.
Instead, we are interested in their correlation.

\begin{figure}
    \centering
    \parbox{\LW{0.1}}{\subcaption{}\label{fig:flowrec3d:params-noisy-selection-notshare}}\hfill%
    \parbox{\LW{0.9}}{\includegraphics[width=0.8\linewidth]{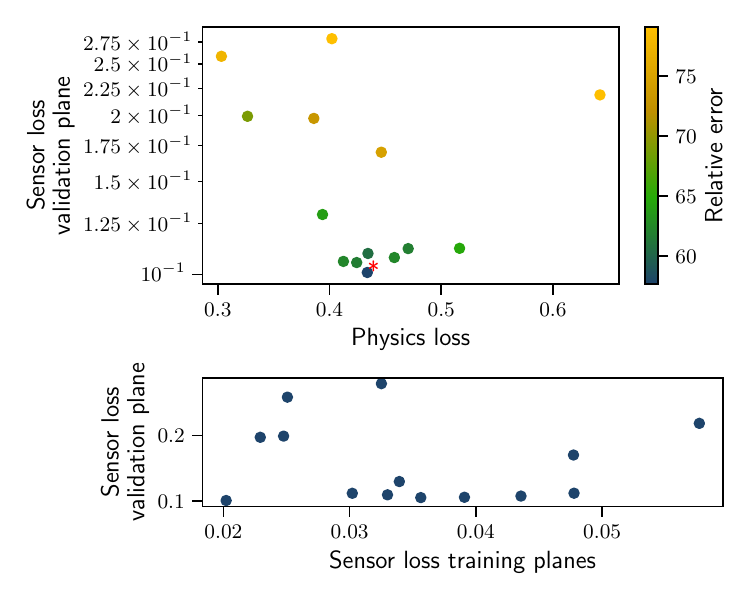}}\hfill%
    \parbox{\LW{0.1}}{\subcaption{}\label{fig:flowrec3d:params-noisy-selection-share}}\hfill
    \parbox{\LW{0.9}}{\includegraphics[width=0.8\linewidth]{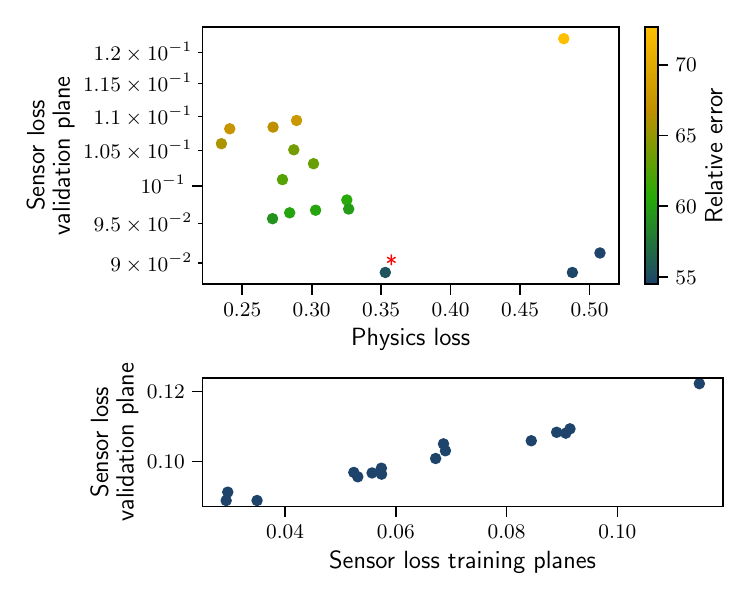}}\hfill%
    \caption{
        Quantities of interest for tests performed with different sets of hyperparameters for (a) the 3D PC-DualConvNet and (b) the weight-sharing network.
        Each data point represents a test with a distinct set of hyperparameters.
        Top panel: the validation sensor loss plotted against the physics loss for tests with different sets of hyperparameters, coloured by the relative error (not used in the selection process). 
        The red star marks the set of hyperparameters selected. 
        Validation sensor loss is computed at the plane at $x_3=$3.14, which is not seen by the networks during training.
        Bottom panel: validation sensor loss against the training sensor loss.
    }\label{fig:flowrec3d:params-noisy-selection}
\end{figure}
Figure~\ref{fig:flowrec3d:params-noisy-selection} shows the quantities of interest for tests performed with different sets of hyperparameters.
The bottom panel of Figure~\ref{fig:flowrec3d:params-noisy-selection-share} shows that the validation sensor loss follows the training sensor loss, showing a linear relationship.
In contrast, no monotonic relationship between the validation and training sensor loss for the PC-DualConvNet is observed (Figure~\ref{fig:flowrec3d:params-noisy-selection-notshare} bottom panel).
By comparing how the validation sensor loss changes with the training sensor loss for the two different network structures, we can see that the weight-sharing network generalises to unseen regions of the flow, which is the purpose of its design.
The linear relationship also means that we can assess the generalisation error of the weight-sharing network by assessing the reconstructed flow on known grid points.
This is not possible with the PC-DualConvNet, as a lower training loss does not correspond to a lower validation loss.   

The top panels of Figure~\ref{fig:flowrec3d:params-noisy-selection-notshare} and~\ref{fig:flowrec3d:params-noisy-selection-share} show how the validation sensor loss changes with the physics loss for PC-DualConvNet and the weight-sharing network, respectively.
The data points are coloured by the relative error.
However, we will not use the relative error during the hyperparameter selection process because the computation of the relative loss requires the full flow field, to which we assume we do not have access.

To select the hyperparameters, we consider two losses: the validation sensor loss and the physics loss.
If the measurements are not noisy, we wish the training process to minimise both losses.
However, in the case of noisy measurements, the lowest validation loss may not correspond to the most accurate reconstruction because the loss is computed with the noisy data $\mat{D}_n$.
Without using information from the ground truth, or the SNR, we also cannot estimate the lower bound for the sensor loss.
Therefore, we also cannot set a threshold for the validation sensor loss.
On the other hand, we cannot choose the set of hyperparameters which leads to the lowest physics loss because a low physics loss only shows that the reconstructed flow is a solution to the Navier-Stokes equation, but does not show whether this solution corresponds to the measurements.
Instead, we look for a compromise by identifying a point where a decrease in the physics loss leads to an increase in the validation sensor loss, and choose the set of hyperparameters at the turning point.
The selected sets of hyperparameters for both networks are marked by a star in Figure~\ref{fig:flowrec3d:params-noisy-selection}, and the values of the selected hyperparameters are listed in Appendix~\ref{sec:app-ml:chap-flowrec3d:params-noisy}.

\subsection{Results from noisy measurements}\label{sec:flowrec3d:results-noisy-results}

Using the hyperparameters selected in Section~\ref{sec:flowrec3d:results-noisy-params}, we reconstruct the 3D turbulent flow from measurement planes shown in Section~\ref{sec:flowrec3d:methods-data} with added white noise at SNR=15.
A summary of the results is shown in Table~\ref{tab:flowrec3d:results-noisy}, where the means and standard deviations are computed over five tests, each with different random initialisation of white noise and network weights.
Similar to the results from non-noisy measurements in Section~\ref{sec:flowrec3d:results-clean}, the weight-sharing network has a lower relative error with a smaller standard deviation, showing that the network becomes less sensitive to the realisation of random noise and weight initialisation by sharing weights.

\begin{figure}
\centering
    \begin{subfigure}{0.9\linewidth}
        \includegraphics[width=\textwidth]{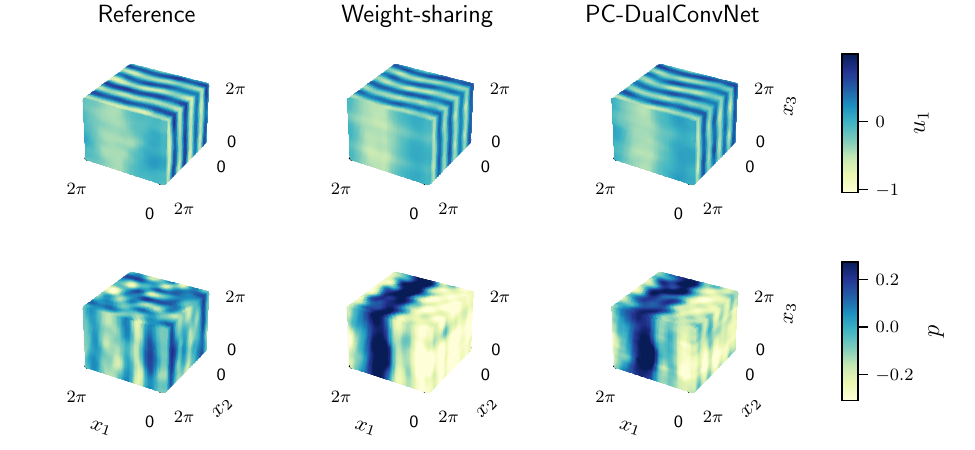}
    \end{subfigure}
    \begin{subfigure}{0.9\linewidth}
        \includegraphics[width=\textwidth]{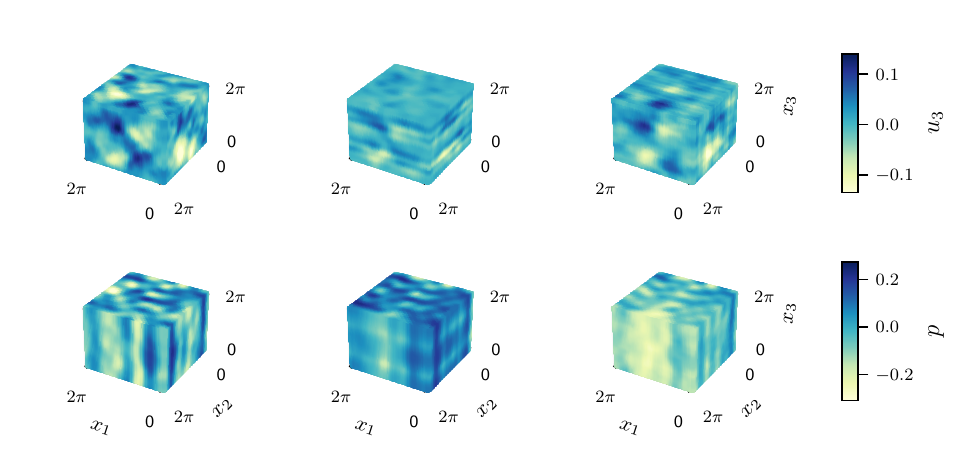}
    \end{subfigure}
    \caption{
        The mean flow. From left to right are: reference 3D data, reconstructed with the weight-sharing network, and reconstructed with the PC-DualConvNet. The top and bottom rows are the vorticity and pressure fields, respectively.
    }\label{fig:flowrec3d:results-noisy-mean}
\end{figure}

Figure~\ref{fig:flowrec3d:results-noisy-mean} shows the time-averaged 3D velocity and pressure fields.
Similar to our observation in Section~\ref{sec:flowrec3d:results-clean}, the two networks perform similarly  when comparing the time-averaged flow on the boundaries of the periodic box, as only the boundaries are visible in Figure~\ref{fig:flowrec3d:results-noisy-mean}.
The reconstructed $u_3$ by the weight-sharing network shows a numerical artefact in the $x_3$ direction, which is the result of the weight sharing.
However, given that the weight-sharing network achieved a similar level of physics loss compared to the PC-DualConvNet, the effect of this artefact is minimal.
\begin{table}
    \caption{
        The relative error and the physics loss (\textit{mean $\pm$ standard deviation}) of the reconstruction results from planes of noisy measurements, averaged over five tests with different random noise and different initialisations of network weights.
    }\label{tab:flowrec3d:results-noisy}
    \centering
    \begin{tabular}{|c|c|c|}
        \hline
         & \textbf{$\epsilon$ (\%)} & \textbf{$\loss_p$} \\
        \hline
        \textbf{Reference data} & N/A & 0.137$\pm$0.000 \\
        \textbf{Weight-sharing network} & 56.7$\pm$0.7 & 1.244$\pm$2.185  \\
        \textbf{PC-DualConvNet} & 59.4$\pm$2.9 & 1.303$\pm$2.156 \\
        \hline
    \end{tabular}
\end{table}

\begin{figure}
    \centering
    \includegraphics[width=0.7\linewidth]{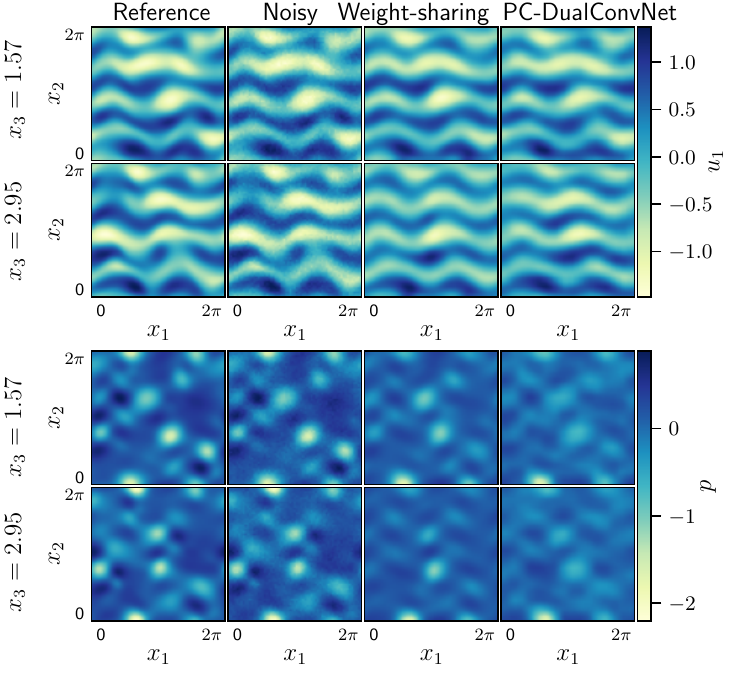}
    \caption{Velocity $u_1$ and pressure $p$ snapshots at different $x_3$. Columns show the reference flow, the flow with added white noise, the flow reconstructed by the weight-sharing network and the flow reconstructed by the PC-DualConvNet (left to right). The top row for each variable shows a plane at $x_3=$1.57, which is part of the training dataset. The bottom row for each variable shows a plane at $x_3=$2.95, which is unseen by the network during training, and also not a validation plane used for selecting the hyperparameters in Section~\ref{sec:flowrec3d:results-noisy-params}.}\label{fig:flowrec3d:results-noisy-different-z}
\end{figure}
At $x_3=$1.57, which is part of the training data, the reconstructed instantaneous $u_1$ for both networks is  less noisy compared to the noisy measurements in the training set (Figure~\ref{fig:flowrec3d:results-noisy-different-z}), and matches well with the reference data.
At the same $x_3$, the reconstructed pressure has a reduced range (the difference between the largest and smallest values is smaller) in the middle of the domain. 
Near $x_2=$0 and 2$\pi$ the pressure reconstruction is more accurate because pressure data is available at $x_2=0$ and periodic boundary conditions are imposed via periodic padding in convolution.
Comparing the slices at $x_3=$2.95, which is unseen in training, we find that the weight-sharing network better captures the main changes in the flow.
Especially in the reconstructed pressure, where the weight-sharing network captured a rotation in the alignment of the two low-pressure areas in the middle of the domain, but not the PC-DualConvNet.
These results show that both networks are capable of reconstructing the flow from noisy measurements, producing reasonable instantaneous flow fields and accurate reconstructed mean flow fields.
The weight-sharing network has proven to be more suitable when the available data is limited to a few planes, as it can generalise to areas of the domain that are away from the measured planes, with fewer parameters.

\FloatBarrier%
\section{Conclusion}\label{sec:flowrec3d:conclusion}

 In this paper, we reconstruct 3D turbulent flows with homogeneity from sparse measurements using a weight-sharing network to infer the full flow field, without relying on ground truth data during training. The measurements comprise three planes of in-flow velocity and one additional plane of boundary pressure. The weight-sharing network applies identical network parameters along the homogeneous direction, enabling more efficient data utilization and reducing computational memory requirements. 
We compare the PC-DualConvNet, adapted from \citet{mo2025ReconstructingUnsteadyFlows}, with the weight-sharing network. First, we reconstruct a 3D Kolmogorov flow from noise-free measurements using the snapshot-enforced loss. Both networks accurately reconstruct time-averaged 3D flow fields and recover the correct energy spectrum up to wavenumber 10, containing most of the flow energy. The weight-sharing network and the PC-DualConvNet achieve relative errors of approximately 49\% and 55\%, respectively. Analysis of reconstructed snapshots shows that the weight-sharing network can infer flow structures in regions distant from the measurement planes.
Second, we reconstruct the flow from measurements with added white noise at a signal-to-noise ratio of 15, using the mean-enforced loss. For the weight-sharing network, we show that the validation sensor loss, which is computed on a plane unseen during training, decreases with the training sensor loss. However, for the PC-DualConvNet, the validation sensor loss does not follow the training sensor loss. Therefore, we conclude that the weight-sharing network generalizes better to unseen regions of the flow, and that the training sensor loss reliably estimates the generalization error for this network. By using the training sensor loss as an estimator, more data can be allocated for training instead of validation, which is beneficial when data is limited. 
The relative errors for flow reconstruction from noisy measurements are approximately 10\% higher than those from noise-free data, but qualitative analysis shows that noise does not significantly impact reconstructed flow structures. In summary, both PC-DualConvNet and the weight-sharing network can reconstruct 3D turbulent Kolmogorov flows from planar measurements, a configuration similar to experimental setups. The weight-sharing network demonstrates good generalization to unseen regions of the flow, minimizes trainable parameters, and offers increased robustness, and less unpredictable behaviours, to hyperparameter selection.

\begin{Backmatter}

\paragraph{Funding Statement}
We acknowledge funding from the Engineering and Physical Sciences Research Council, UK and financial support from the ERC Starting Grant PhyCo 949388. L.M. is also grateful for the support from the grant EU-PNRR YoungResearcher TWIN ERC-PI\_0000005. 

\paragraph{Competing Interests}
The authors declare no conflict of interest. 

\paragraph{Data Availability Statement}
The codes to perform all the tests in this paper can be found at \url{https://github.com/MagriLab/FlowReconstructionFromExperiment}.
All data is available upon request.

\paragraph{Ethical Standards}
The research meets all ethical guidelines, including adherence to the legal requirements of the study country.

\paragraph{Author Contributions}
All authors have read and approved the final manuscript.

\bibliographystyle{apalike}
\bibliography{reference.bib}

\end{Backmatter}

\FloatBarrier%
\begin{appendix}
\section*{Appendices}
\input{appendix}
\end{appendix}

\end{document}

%% file: appendix.tex
\section{Hyperparameters and the minimum number of planes}\label{sec:app-ml:chap-flowrec3d-params-and-min-planes}
\begin{figure}
    \centering
    \includegraphics[width=0.8\linewidth]{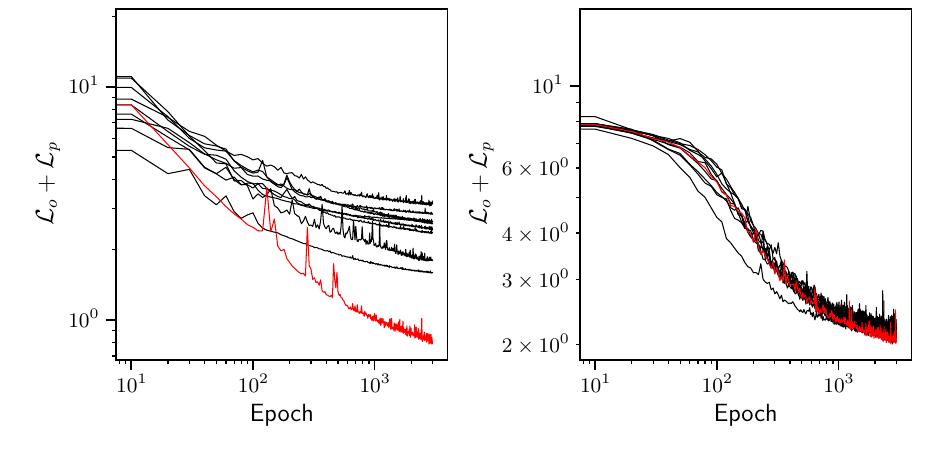}
    \caption{The unweighted sum of the physics and sensor loss of the tests in the hyperparameter selection. Left: PC-DualConvNet. Right: weight-sharing network.}\label{fig:app-ml:chap-flowrec3d:sweep-reduce-planes}
\end{figure}

\begin{figure}
    \centering
    \includegraphics[width=0.9\linewidth]{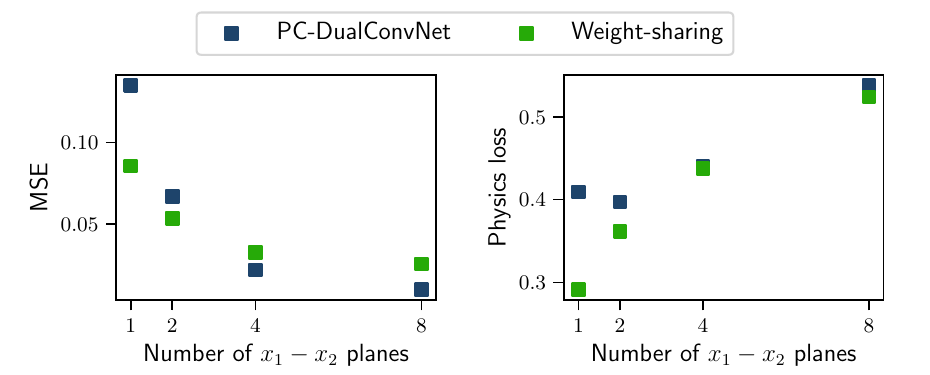}
    \caption{(Left) the mean squared error and (right) the physics loss of the reconstructed from fields from different number of $x_1-x_2$ measurement planes.}\label{fig:app-ml:chap-flowrec3d:reduce-planes}
\end{figure}

In this section, we test the minimum number of $x_1-x_2$ measurement planes needed to reconstruct the 3D Kolmogorov flows using both the PC-DualConvNet and the weight-sharing network. We reduce the number of $x_1-x_2$ measurement planes, while keeping the single $x_2-x_3$ plane unchanged, until the relative errors of the reconstructed flows from both networks exceed 50\%. We start by selecting a set of hyperparameters to reduce the unweighted sum of the physics and sensor loss ($\loss_p + \loss_o$) using test cases with eight $x_1-x_2$ measurement planes, evenly spaced in the $x_3$ direction.
Figure~\ref{fig:app-ml:chap-flowrec3d:sweep-reduce-planes} shows the training curves of the tests conducted during the hyperparameter selection process; each test uses a different set of hyperparameters. The tests with the selected hyperparameters are highlighted in red. Figure~\ref{fig:app-ml:chap-flowrec3d:sweep-reduce-planes} (left panel) shows the tests using PC-DualConvNet. Among the unselected tests (black), the differences in hyperparameters result in different final loss values, but the training curves follow a similar downward trend. However, the test with the selected hyperparameters (red) clearly shows a faster decrease of the loss than the other tests (black). This difference in the training curve highlights the importance of hyperparameters in the training of the PC-DualConvNet.
The selected hyperparameters are detailed in Table~\ref{tab:app-ml:chap-flowrec3d:clean-params-PC-DualConvNet} and~\ref{tab:app-ml:chap-flowrec3d:clean-params-sharing}. 
The learning rate schedules are taken from \citep{mo2025ReconstructingUnsteadyFlows}.
\begin{table}
    \centering
    \caption{
        The hyperparameters of the PC-DualConvNet used to reconstruct the 3D Kolmogorov flows in Section~\ref{sec:flowrec3d:results-clean}.
    }\label{tab:app-ml:chap-flowrec3d:clean-params-PC-DualConvNet}
    \begin{tabular}{|c|c|}
        \hline
        Bottleneck image dimension & (4,4,4) \\
        Convolution filter size (all) & (3,3,3) \\
        Convolution layer padding & Periodic \\
        Input branch channels & [4,] \\
        Upper branch channels & [4,] \\
        Lower branch channels & [4,8,8,4] \\
        Output branch channels & [4,] \\
        FFT & off \\
        Batch size & 250 \\
        $\mathbf{\lambda}_{div}$ & 1.0 \\
        $\mathbf{\lambda}_{mom}$ & 3.0 \\
        Initial learning rate ($\alpha$) & 0.001 \\
        Learning rate schedule & Cyclic decay \\
        regularization & 0.0 \\
        Dropout rate & 0.0 \\
        \hline
    \end{tabular}
\end{table}
\begin{table}
    \centering
    \caption{
        The hyperparameters of the weight-sharing network used to reconstruct the 3D Kolmogorov flows in Section~\ref{sec:flowrec3d:results-clean}.
    }\label{tab:app-ml:chap-flowrec3d:clean-params-sharing}
    \begin{tabular}{|c|c|}
        \hline
        Bottleneck image dimension (2D) & (8,8) \\
        Convolution filter size (2D) & (3,3,3) \\
        Upper branch channels (2D) & [4,] \\
        Lower branch channels (2D) & [4,16,16,8] \\
        FFT (2D) & off \\
        Bottleneck image dimension (3D) & (32,32,32) \\
        Channels (3D) & [8,8,4] \\
        Convolution filter size (3D) & [(3,3,3),(5,5,5),(5,5,5)] \\
        Convolution layer padding (all) & Periodic \\
        Batch size & 100 \\
        $\mathbf{\lambda}_{div}$ & 1.0 \\
        $\mathbf{\lambda}_{mom}$ & 2.0 \\
        Initial learning rate ($\alpha$) & 0.0025 \\
        Learning rate schedule & Exponential decay \\
        regularization & 0.0 \\
        Dropout rate & 0.0 \\
        \hline
    \end{tabular}
\end{table}
 Using the selected set of hyperparameters, we reduce the number of measurement planes and plot the resulting mean squared error (MSE) and physics loss of the reconstructed flow fields in Figure~\ref{fig:app-ml:chap-flowrec3d:reduce-planes}. By MSE, we refer to the $\ell_2$ norm of the difference between the reference and reconstructed flow at all grid points, $\| \mat{D} - \hat{\mat{D}} \|^2_2$. Both the MSE and the physics loss show that the weight-sharing network achieves lower values for both metrics when there are two or fewer $x_1-x_2$ measurement planes.
The relative errors from both networks when there is only one $x_1-x_2$ plane are much larger than 50\%, and the reconstructed flow fields lose resemblance to the reference data. Therefore, we report the results of reconstructing the 3D Kolmogorov flows from two $x_1-x_2$ planes in the main text (Section~\ref{sec:flowrec3d:results-clean}).
\FloatBarrier%
\section{Model training time}\label{sec:app-ml:chap-flowrec3d:timer}
This section discusses the training time of the networks used in Section~\ref{sec:flowrec3d:results-clean}. Table~\ref{tab:app-ml:chap-flowrec3d:update-time} shows the number of parameters and the average update and inference time of the PC-DualConvNet and the weight-sharing network. The PC-DualConvNet has approximately 500 times more parameters than the weight-sharing network, and approximately double the inference time, meaning that the weight-sharing network uses less memory and is faster during inference. However, the average update times of the two networks are within 10\% of each other, which means that both networks take a similar amount of time to train despite the large difference in the number of parameters. Upon closer inspection, we found the bottleneck of the update time to be the computation of the physics loss.
Table~\ref{tab:app-ml:chap-flowrec3d:compute-loss-time} shows that the average time to compute the physics loss for a 50-snapshot batch is 84.4~ms, which is over 40\% of the update times for both networks, and over double the time taken to compute the mean squared error (MSE) of the full flow field. The time to compute the sensor loss is negligible in comparison. When scaling up the datasets for future work, we must take into account the time to compute the physics loss and investigate ways to speed up this computation.

\begin{table}[hbt]
    \centering
    \caption{Number of trainable parameters; the average time to compute one update (including a forward pass, computing the loss, and applying the update); and the average inference time (forward pass only). Timed using a 50-snapshot batch on an NVIDIA RTX8000 GPU.\@ Functions are compiled with \lstinline{jax.jit}.}\label{tab:app-ml:chap-flowrec3d:update-time}
    \begin{tabular}{|p{4cm}|c|c|c|}
        \hline
         & \textbf{Num.\ of params} & \textbf{Update time} (ms) & \textbf{Inference time} (ms) \\
        \hline
        \textbf{Weight-sharing} & 271,805 & 208 & 46.4 \\
        \textbf{PC-DualConvNet} & 134,255,824 & 189 & 107 \\
        \hline
    \end{tabular}
\end{table}
\begin{table}[hbt]
    \centering
    \caption{The average time to compute loss using a 50-snapshot batch on an NVIDIA RTX8000 GPU.\@ Functions are compiled with \lstinline{jax.jit}.}\label{tab:app-ml:chap-flowrec3d:compute-loss-time}
    \begin{tabular}{|c|c|}
        \hline
         & \textbf{Computation time} (ms) \\
        \hline
        \textbf{Physics loss} & 84.4  \\
        \textbf{Sensor loss} & 2.2 \\
        \textbf{MSE of full field} & 38.8 \\
        \hline
    \end{tabular}
\end{table}

\section{Model and training parameters for noisy data}\label{sec:app-ml:chap-flowrec3d:params-noisy}
Table~\ref{tab:app-ml:chap-flowrec3d:noisy-params-PC-DualConvNet} and~\ref{tab:app-ml:chap-flowrec3d:noisy-params-sharing} show the hyperparameters used to reconstruct the 3D Kolmogorov flows from noisy measurements in Section~\ref{sec:flowrec3d:results-noisy} using the PC-DualConvNet and the weight-sharing network, respectively.
\begin{table}
    \centering
    \caption{
        The hyperparameters of the PC-DualConvNet used to reconstruct the 3D Kolmogorov flows in Section~\ref{sec:flowrec3d:results-noisy}.
    }\label{tab:app-ml:chap-flowrec3d:noisy-params-PC-DualConvNet}
    \begin{tabular}{|c|c|}
        \hline
        Bottleneck image dimension & (4,4,4) \\
        Convolution filter size (all) & (3,3,3) \\
        Convolution layer padding & Periodic \\
        Input branch channels & [4,] \\
        Upper branch channels & [4,] \\
        Lower branch channels & [4,8,8,4] \\
        Output branch channels & [4,] \\
        FFT & off \\
        Batch size & 50 \\
        $\mathbf{\lambda}_{div}$ & 1.0 \\
        $\mathbf{\lambda}_{mom}$ & 4.0 \\
        $\mathbf{\lambda}_{o}$ & 32.0 \\
        Initial learning rate ($\alpha$) & 0.0043 \\
        Learning rate schedule & Cyclic decay \\
        regularization & 0.0 \\
        Dropout rate & 0.0094 \\
        \hline
    \end{tabular}
\end{table}
\begin{table}
    \centering
    \begin{tabular}{|c|c|}
        \hline
        Bottleneck image dimension (2D) & (8,8) \\
        Convolution filter size (2D) & (3,3,3) \\
        Upper branch channels (2D) & [4,] \\
        Lower branch channels (2D) & [4,16,16,8] \\
        FFT (2D) & off \\
        Bottleneck image dimension (3D) & (32,32,32) \\
        Channels (3D) & [8,8,4] \\
        Convolution filter size (3D) & [(3,3,3),(5,5,5),(5,5,5)] \\
        Convolution layer padding (all) & Periodic \\
        Batch size & 50 \\
        $\mathbf{\lambda}_{div}$ & 1.0 \\
        $\mathbf{\lambda}_{mom}$ & 14.0 \\
        $\mathbf{\lambda}_{o}$ & 45.0 \\
        Initial learning rate ($\alpha$) & 0.0026 \\
        Learning rate schedule & Cyclic decay \\
        regularization & 0.0028 \\
        Dropout rate & 0.004 \\
        \hline
    \end{tabular}
    \caption{
        The hyperparameters of the weight-sharing network used to reconstruct the 3D Kolmogorov flows in Section~\ref{sec:flowrec3d:results-noisy}.
    }\label{tab:app-ml:chap-flowrec3d:noisy-params-sharing}
\end{table}